# Intelligent Adaptive Metasurface in Complex Wireless Environments

Han Qing Yang [†], Jun Yan Dai [†], Hui Dong Li [†], Lijie Wu, Meng Zhen Zhang, Zi Hang Shen, Si Ran Wang, Zheng Xing Wang, Wankai Tang, Shi Jin, Jun Wei Wu, Qiang Cheng[*], Tie Jun Cui[*]

[†] Equally contributed to this work.
[*] Email: qiangcheng@seu.edu.cn, tjcui@seu.edu.cn

**Abstract**

The programmable metasurface is regarded as one of the most promising transformative technologies for next-generation wireless system applications, due to Due to the lack of effective perception ability of the external electromagnetic environment, there are numerous challenges in the intelligent regulation of wireless channels, and it still relies on external sensors to reshape electromagnetic environment as desired. To address that problem, we propose an adaptive metasurface (AMS) which integrates the capabilities of acquiring wireless environment information and manipulating reflected electromagnetic (EM) waves in a programmable manner. The proposed design endows the metasurfaces with excellent capabilities to sense the complex electromagnetic field distributions around them and then dynamically manipulate the waves and signals in real time under the guidance of the sensed information, eliminating the need for prior knowledge or external inputs about the wireless environment. For verification, a 6×6 prototype of the proposed AMS is constructed, and its dual capabilities of sensing and manipulation are experimentally validated. Additionally, different integrated sensing and communication (ISAC) scenarios with and without the aid of the AMS are established. The effectiveness of the AMS in enhancing communication quality is well demonstrated in complex electromagnetic environments, highlighting its beneficial application potential in future wireless systems.

**Introduction**

The performances of modern radio systems are intricately tied to a host of essential factors like the transmitting power, the antenna characteristics as well as the receiver sensitivity. In addition, the propagation environment of electromagnetic (EM) waves also exerts a determining influence on the system efficacy. For instance, in open spaces such as deserts or plains, it is easy for wireless signals to travel over long distances with relatively small attenuation. However, in complex environments such as urban areas with densely packed high-rise buildings and moving vehicles, the spread of EM waves will encounter various physical processes, including multiple reflections, refractions, and scatterings, leading to severe signal fading and distortions that significantly degrade the system quality. The unpredictability of wave propagation introduces variability in signal quality and reliability across various applications, such as communication, radar, and positioning, posing significant challenges to the overall effectiveness of the radio systems [1-3]. Consequently, substantial efforts have been made to acquire sufficient information about the propagation environment to mitigate the risk of system failures and ensure reliable operation [4-7]. For instance, to improve wireless communication quality, people are actively striving to acquire channel state information (CSI) through diverse methods [40-41], making it possible to dynamically alter the transmission parameters and error correction methods for reliable and efficient data conveyance.

Recent advances of the information metasurface furnish a fresh approach for free manipulation of wireless propagation environments [8-15]. By describing various reflection/transmission states of the element with binary coding, the information metasurface exhibits prominent capabilities over the control of EM wave properties (amplitude, phase, polarization, etc.) in a digital way, allowing for a plenty of applications like beam shaping, spectral modulation, polarization conversion and vortex wave generation based on different space-time coding strategies [16-21]. Furthermore, the information metasurface can be reconfigured in real-time and switched among various functionalities, which renders an excellent hardware platform to confront various scenarios for wireless applications in complicated propagation environments [22-39].

However, despite the formidable performances of wave manipulation for the information metasurfaces, they still lack effective sensing means to secure wireless channel information in order to supply a ground for efficient wave control strategies. Thus, it is hard to reshape the wireless propagation environment effectively and raise the system's performance to the greatest possible extent. To tackle this challenge, more and more attempts have been made to endow the metasurface with the power of sensing, that spurred the generation of adaptive metasurfaces (AMS). One route to realize the AMS is that the metasurface reflects most of the impinging signals, while a small fraction is coupled into sensing chains embedded within each element [5-6, 42-47]. Another involves incorporating active modules (such as antennas) into certain elements to enable direct reception of the incident signals for sensing [7]. An additional option is to rest on external sensors to present auxiliary sensing information to traditional metamaterials [48].

Nevertheless, the present solutions still encounter numerous bottleneck issues. Firstly, most reported AMSs can only acquire the magnitude of the incident signal without phase information. Secondly, the phase quantization bit of the AMS is limited to 1- or 2-bit, as will produce slight influence on the precision of the scattering beam pointing. Thirdly, the existing AMSs are deficient in sensing and regulation capabilities at the unit level. It dramatically lowers the flexibility of the metasurfaces for efficient wave control.

Confronted with the aforementioned challenges, in this paper we have proffered a new type of intelligent AMS with enhanced capabilities. The proposed AMS can simultaneously acquire both the magnitudes and the phases of the surrounding EM waves. This considerably improves the sensing power of AMS, making it possible to acquire information from a broader perspective and swiftly modify its own scattering features to adapt to complex propagation environments. The metasurface also confers the ability of continuous phase alteration with full phase range, while maintaining the reflection magnitude higher than 0.84. This is especially valued to promote the accuracy of scattering beam control. Furthermore, the proposed design employs a reception-reflection scheme that facilitates element-by-element sensing and manipulating. To demonstrate the powerful ability of the metasurface, three representative integrated sensing and communication (ISAC) scenarios are considered, and the performances of AMS are experimentally validated. The results display good application vistas of the AMSs in boosting the performances of current wireless systems.

**Methods**

**Fig. 1** illustrates the schematic diagram of the wireless environment regulation empowered by the proposed AMS and a closed - loop working paradigm involving sensing, analyzing, and manipulating. The AMS can sense the incident field element - by - element across its entire surface and obtain extensive wireless environmental information. After obtaining the environmental information, analysis is carried out to obtain the reflection matrix $\mathbf{\Phi}$ of the elements by an optimization algorithm, which is employed to characterize the element phases in order to obtain the desired EM field distributions at the surface. The microcontroller unit (MCU) then generates DC control signals to adjust the working states of all AMS elements corresponding to the matrix $\mathbf{\Phi}$.

To illustrate the working principle of AMS, **Fig. 2a** portrays a typical scenario of wireless environment regulation, where the sources, the scatters, and the metasurface are all present simultaneously. $\mathbf{J}(\mathbf{r})$ and $\mathbf{M}(\mathbf{r})$ are equivalent electric and magnetic currents at the position $\mathbf{r} = [r_x, r_y, r_z]$ as the EM radiators. $\varepsilon(\mathbf{r})$, $\mu(\mathbf{r})$ and $\sigma(\mathbf{r})$ stand for the permittivity, permeability, and conductivity at different locations in space respectively, which can clearly describe the material properties of various complex scatterers in the environment. The AMS has $M \times N$ elements that can dynamically respond to the incident EM waves. The propagation and manipulation processes can be described by Equation (1) (see details in Supplementary Information)

$$\begin{aligned}\mathbf{E}(\mathbf{r}) &= \mathbf{E}_{in}(\mathbf{r}) + \sum_{m=1}^{M}\sum_{n=1}^{N}(-j\omega\mu_0)\left[1 + \frac{1}{k^2}\nabla\nabla\right]\mathbf{J}^s(\mathbf{r}_{m,n}) \cdot G(\mathbf{r},\ \mathbf{r}_{m,n}) \\ &= \mathbf{E}_{in}(\mathbf{r}) + \sum_{m=1}^{M}\sum_{n=1}^{N}(-j\omega\mu_0)\left[1 + \frac{1}{k^2}\nabla\nabla\right] \cdot \Gamma_{m,n} \cdot \mathbf{Z}^H \cdot \mathbf{E}_{in}(\mathbf{r}_{m,n}) \cdot G(\mathbf{r},\ \mathbf{r}_{m,n})\end{aligned} \quad (1)$$

Here $\mathbf{E}(\mathbf{r})$ and $\mathbf{E}_{in}(\mathbf{r})$ denote the total and incident electric field evaluated at the point **r**, respectively. $\mathbf{r}_{m,n}$ and $\Gamma_{m,n} = A_{m,n} e^{j\phi_{m,n}}$ are the location and the complex reflection coefficient of the AMS element $(m, n)$. $G(\mathbf{r}, \mathbf{r}_{m,n})$ is the Green's function of the scalar Helmholtz equation, which is closely related to the wireless environment. The Green's function between two points **r** and **r**′ can be expressed as $G(\mathbf{r}, \mathbf{r}') = -e^{-jk|\mathbf{r}-\mathbf{r}'|}/4\pi|\mathbf{r}-\mathbf{r}'|$ in free space. However, regarding a complex EM circumstance populated with irregular scatterers, it is extremely tough to predict or portray the Green's function analytically.

For conventional EM metasurface, it is possible to change the reflection matrix **Φ** for wave manipulation with the help of optimization methods [27-32], where
$$\mathbf{\Phi} = \begin{bmatrix} \Gamma_{1,1} & \Gamma_{1,2} & \cdots & \Gamma_{1,N} \\ \Gamma_{2,1} & \Gamma_{2,2} & \cdots & \Gamma_{2,N} \\ \vdots & \vdots & \ddots & \vdots \\ \Gamma_{M,1} & \Gamma_{M,2} & \cdots & \Gamma_{M,N} \end{bmatrix} \in \mathbb{C}^{M \times N}$$
. However, in complex wireless environments, the Green's function $G$ in equation (1) is unknown and unpredictable. As a result, the matrix **Φ** cannot always be guaranteed to be the most optimal and may not be the best fit for the channel. An improved way is to directly get the channel information as the basis for adjusting the reflection matrix. But this requires the addition of extra sensors or antennas. And these devices will share an aperture with the metasurface units, which not only increases the cost and complexity but also has a non-negligible influence on the EM performance of the metasurface. It is therefore suggested that if the metasurface can directly access information from the incident waves in an element-by-element way, it would significantly enhance sensing accuracy and real-time performance while minimizing the cost.

To handle the aforementioned issue, this study introduces a new design for AMS. Each element is able to directly acquire information from the environment, thus establishing a new closed-loop working paradigm for sensing, analyzing, and manipulating. It combines autonomous analysis and decision without relying on pre-predicted information from external sensors. This greatly improves the flexibility of the metasurface for EM manipulation, minimizes the impact on the overall reflection performance owing to the improvement of the sensing ability, and enhances the real-time response speed of the system.

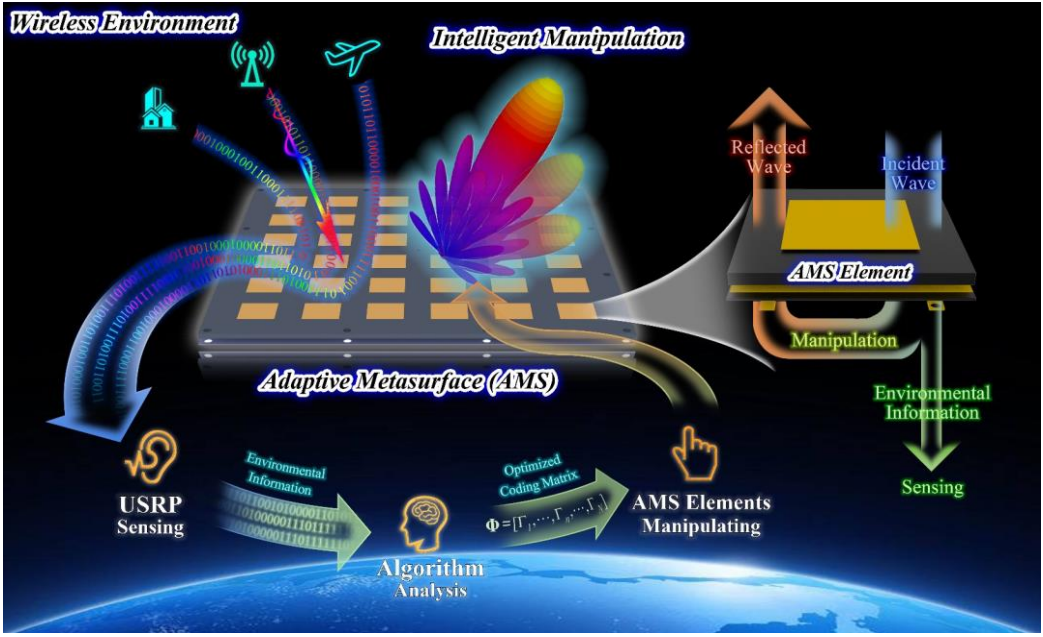

**Fig. 1.** Adaptive metasurface (AMS) for intelligent wireless environment regulation.

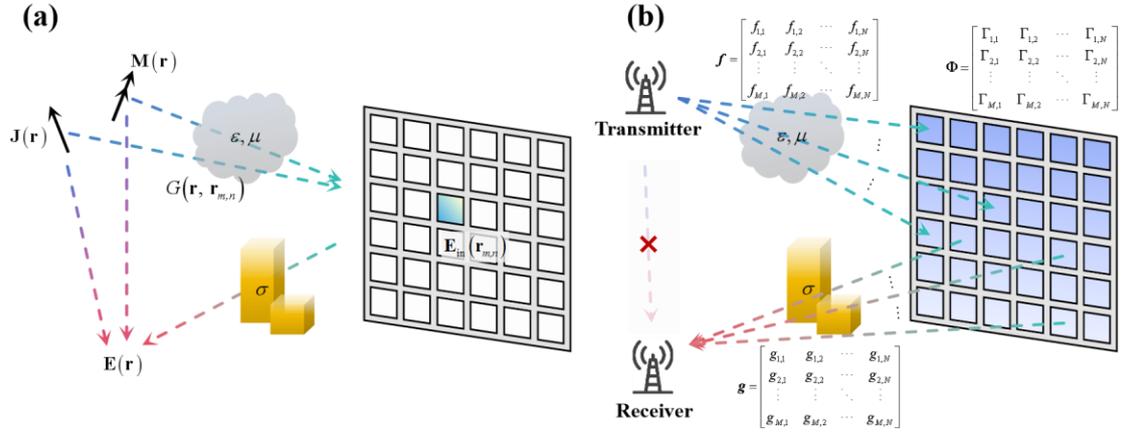

**Fig. 2**. **(a)** Wireless environment regulation with the aid of traditional metasurface, including various sources and scatters. **(b)** Communication scenario in the regulated wireless environment based on the proposed AMS; *f* and *g* are matrices describing the channel state information; **Φ** is AMS's reflection matrix.

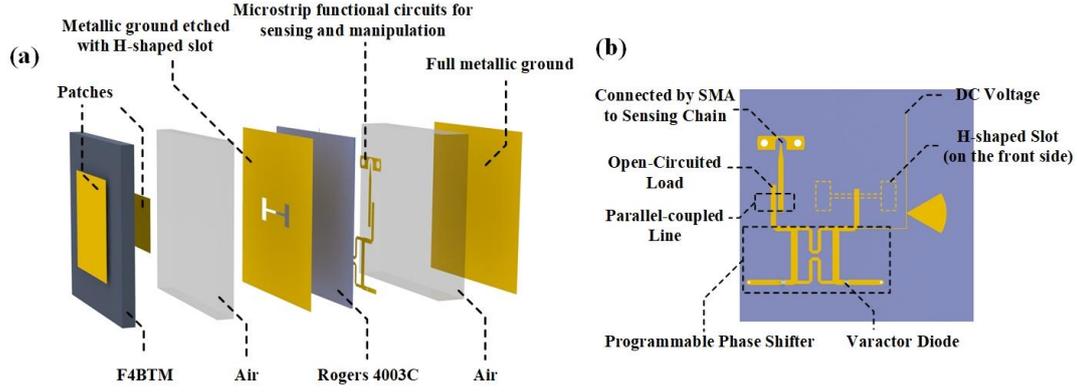

**Fig. 3**. (**a**) The structure of the proposed AMS element. (**b**) Details of the sensing and manipulating circuit. Geometric details of the proposed element can be found in Supplementary Information.

**AMS element design**

The detailed design of the proposed AMS can be found in **Fig. 3a**. The element has a size of 40×40 mm$^2$ (0.47×0.47 wavelengths at 3.50 GHz) and comprises three layers. The top layer consists of an F4BTM substrate sandwiched by two metallic patches with different dimensions. The middle layer is made of a Rogers 4003C substrate, whose front side has a metallic ground etched with an H - shaped slot, while the reverse has microstrip functional circuits for sensing and manipulation. The bottom layer is a full metallic ground. Note that the two patches and the H-shaped slot amalgamate to form an antenna, which cooperates with the microstrip circuit to couple in a portion of the energy of the incident EM wave. The bottom metallic ground is employed to enhance the reflection efficiency of the AMS (See more details in Supplementary Information).

The manipulating and sensing principles of the proposed AMS are conceptually summarized as follows. Initially, the incident wave is received by the slot antenna and subsequently coupled to the sensing and manipulating circuits (refer to **Fig. 3b**).

Then the received signal firstly passes through a phase shifter. The programmable phase shift is achieved through different equivalent capacitances $C_{VAR}$ of the internal varactor diodes by applying an external DC voltage $V_{DC}$. After the phase shifter, the signal then reaches a weakly parallel-coupled line. Here, a small fraction of the signal energy is coupled to the parallel microstrip line for sensing, while the majority continues to propagate forward until being reflected by the open-circuited load. Finally, the reflected signal follows its original path back. It is adjusted by the phase shifter again and sent back into free space by the same slot antenna. Meanwhile, the signal coupled into the parallel microstrip line enters the sensing chain and helps in generating the reflection matrix **Φ** for the AMS.

The manipulating and sensing capabilities of the AMS elements are first evaluated through full-wave simulation. To evaluate the manipulating performance, the reflection coefficient of the AMS element is defined as the ratio $\Gamma = \mathbf{E}_{out}/\mathbf{E}_{in}$, in which $\mathbf{E}_{out}$ and $\mathbf{E}_{in}$ represent the complex amplitudes of the reflected and incident wave, respectively. The relationship between Γ and $C_{VAR}$ is depicted in **Figs. 4a** and **4b**. As $C_{VAR}$ varies between 0.140 pF to 0.710 pF, the phase of Γ continuously varies within the range of 0° to 360°, demonstrating the large phase tuning range of the proposed AMS element. To assess the EM sensing performance, the coupling coefficient of the AMS element is defined as the ratio $\alpha = \mathbf{E}_{cou}/\mathbf{E}_{in}$, in which $\mathbf{E}_{cou}$ represents the complex amplitude of the signal coupled into the chain. The dependence of α on $C_{VAR}$ is depicted in **Figs. 4g and 4h**. It can be observed that the presented AMS element can provide a stable and moderate coupling at the desired frequency for EM sensing. It is worth noting that the absolute value of α is not crucial, because the primary concern often lies in the relative magnitude or phase values of the environmental parameters in wireless systems.

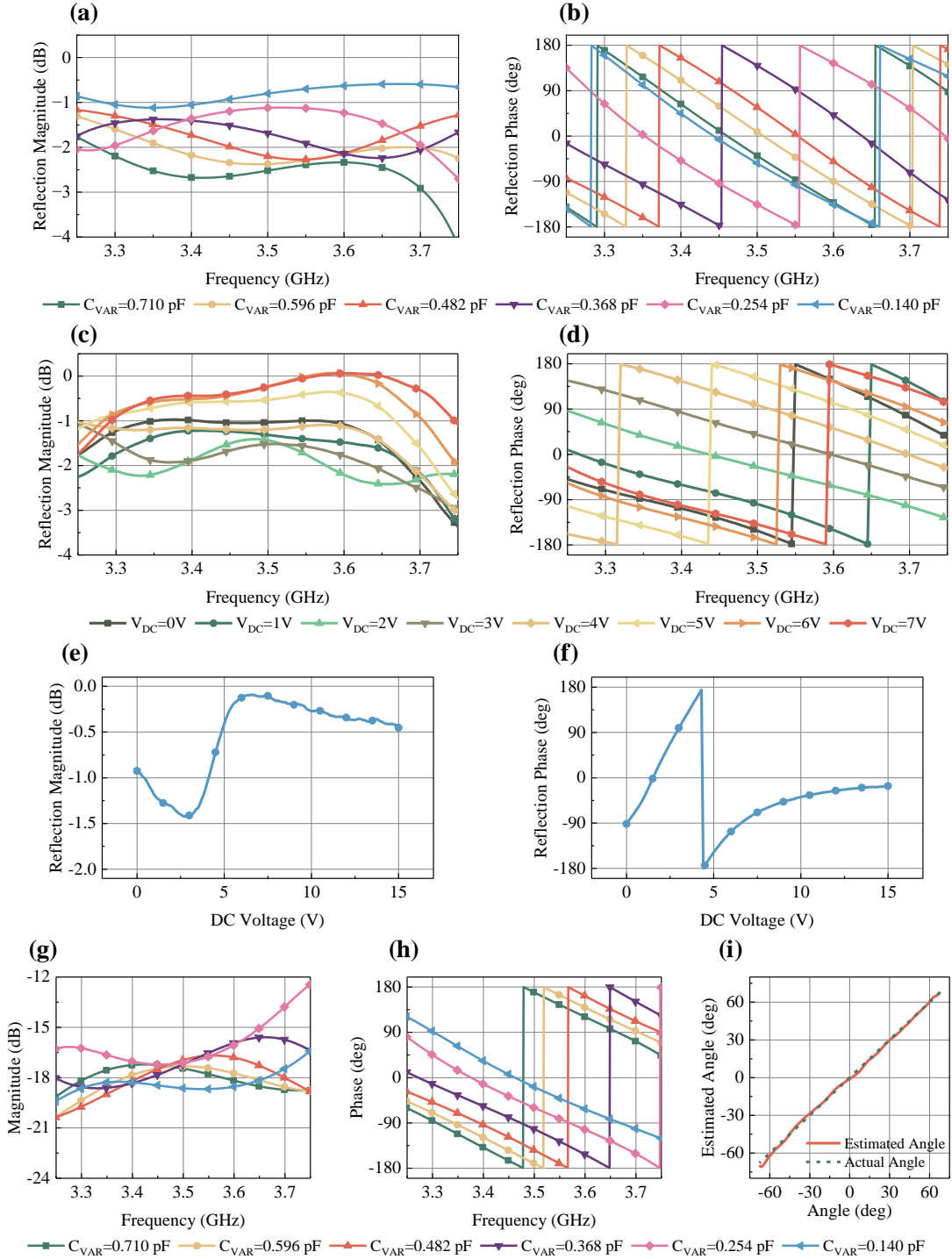

**Fig. 4.** **(a, b)** Dependence of the simulated reflection magnitude and phase spectra on $C_{VAR}$. **(c, d)** Dependence of the measured reflection magnitude and phase spectra on $V_{DC}$. **(e, f)** Measured reflection magnitude and phase spectra at 3.50 GHz as $V_{DC}$ varies from 0 to 15V. **(g, h)** Dependence of the simulated coupling magnitude and phase on $C_{VAR}$. **(g)** The relationship between the actual angle and the estimated angle in the DOA estimation experiment with the proposed AMS.

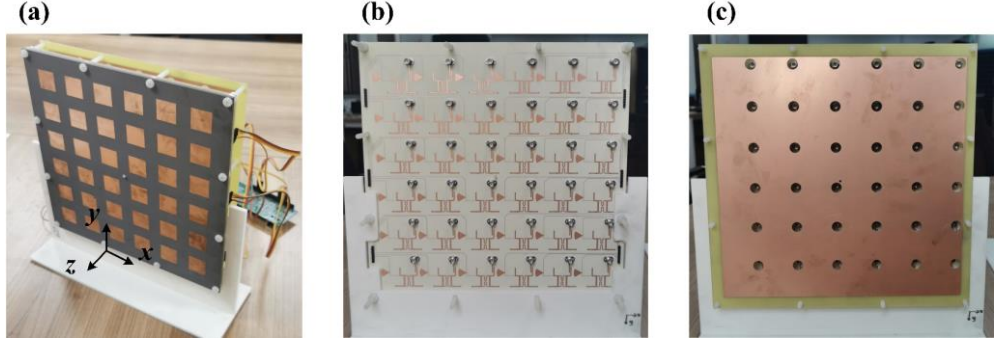

**Fig. 5**. (**a**) Prototype of AMS comprising 6×6 elements. (**b, c**) The back view of the prototype with and without the bottom metallic ground. The circular holes on the ground serve as interfaces to the sensing chains.

**Experimental Validation**

In this section, the capabilities of EM manipulating and sensing based on the proposed AMS are experimentally verified. A prototype of the proposed AMS comprising 6×6 identical elements is fabricated for validation. There are 36 individual DC voltages supplied by an MCU to control the elements independently, as shown in **Fig. 5**. The relationship between the reflection coefficient $\Gamma$ and the biasing voltage $V_{DC}$ for each element is extracted and depicted in **Figs. 4c to 4f**. As $V_{DC}$ is gradually raised from 0 to 7V, the reflection phase of $\Gamma$ can be continuously tuned in the range of 0° to 360°, while its magnitude remains above 0.84 (-1.45dB). The relationship between $C_{VAR}$ in the simulation and $V_{DC}$ in the experiment can be found in the phase shifter section of the Supplementary Information.

Initially, we experimentally confirm the AMS's capacity to achieve dynamic beaming in a programmable fashion, with the traditional beamforming algorithm given in Equation (S4) (in Supplementary Information). The measurement setup is shown in **Fig. 6a**. The AMS prototype is placed on a turntable in the anechoic chamber. A horn antenna serves as the source, with the incident wave direction oriented at $\theta = 0°$ and -20° respectively in the plane of $\varphi = 0°$. In the far-field region, another receiving horn antenna is employed to record the scattering pattern. The measurement results, depicted in **Figs. 7a** and **7b**, indicate that the scattering beam can scan within a range of ±60°.

It is clear that there are slight beam distortion near 0° and -20° stemming from the obstruction of the antenna in front of the metasurface.

. Then we proceed to assess the sensing capability of the AMS. It is noteworthy that all AMS elements are furnished with sensing chains, thereby endowing each element with the capability to capture both the magnitude and phase information of the incident wave. In the experiment, a source is positioned in the far-field zone, and the emitted electromagnetic waves are incident upon the AMS obliquely. The received magnitudes and phases by the 6×6 sensing elements are respectively presented in Figs. 7d and 7f. In the meanwhile, the theoretical field distributions on the AMS in free space can be obtained from Equation (S1) in the Supplementary Information, as shown in Figs. 7c and 7e. Given that the propagation environment in the microwave chamber is close to free space, the experimental results are in good accordance with the theoretical predictions. Additionally, taking the amplitude and phase information collected by each element as a reference, the proposed AMS can easily complete the work of DOA estimation. (see more details in the Supplementary Information). From Fig. 4i, the measured results demonstrate a small estimation error below 1° when the incident angle is within ±60°. Before the experiment, the 6×6 sensing chains are connected to an RF switch matrix. During the measurements, the switch matrix is electronically controlled to quickly toggle among selected links, allowing for rapid information acquisition from all 36 signal paths. Further details on the RF switch matrix can be found in the Supplementary Information.

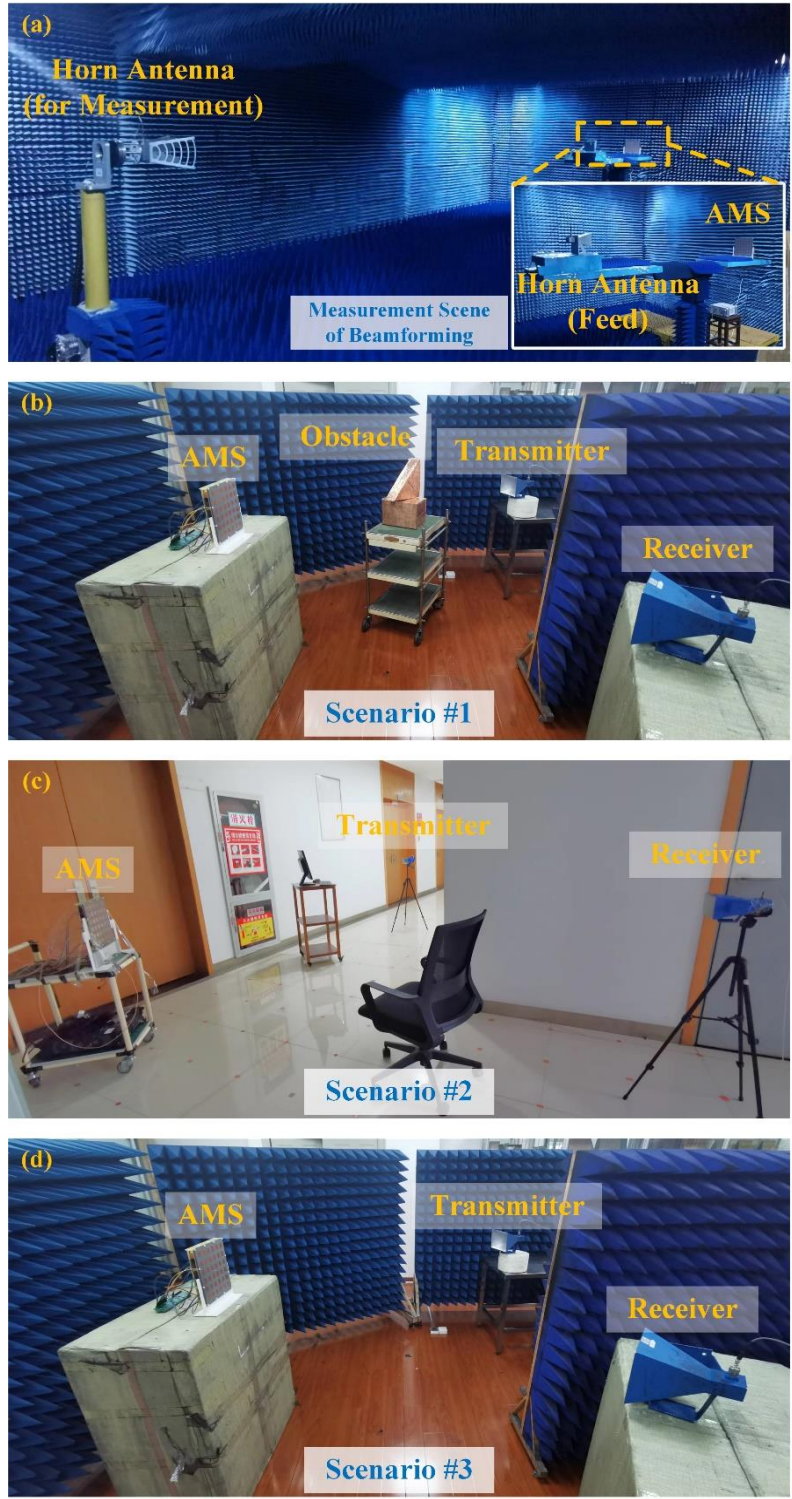

**Fig. 6.** (**a**) Measurement configurations of the dynamic beamforming. (**b-d**) Measurement configurations of three ISAC scenarios. **Scenario #1**: The signal is transmitted non-line-of-sight between the transmitter and the receiver through the metasurface, while the propagation path is blocked by a triangular metallic obstacle. The surrounding absorbing materials are used to eliminate the multipath effect. **Scenario #2**: The signal is transmitted between the transmitter and the receiver in an open corridor environment with distinct multipath effect. **Scenario #3**: The signal is transmitted non-line-of-sight between the transmitter and the receiver through the metasurface without any obstacle in the propagation path.

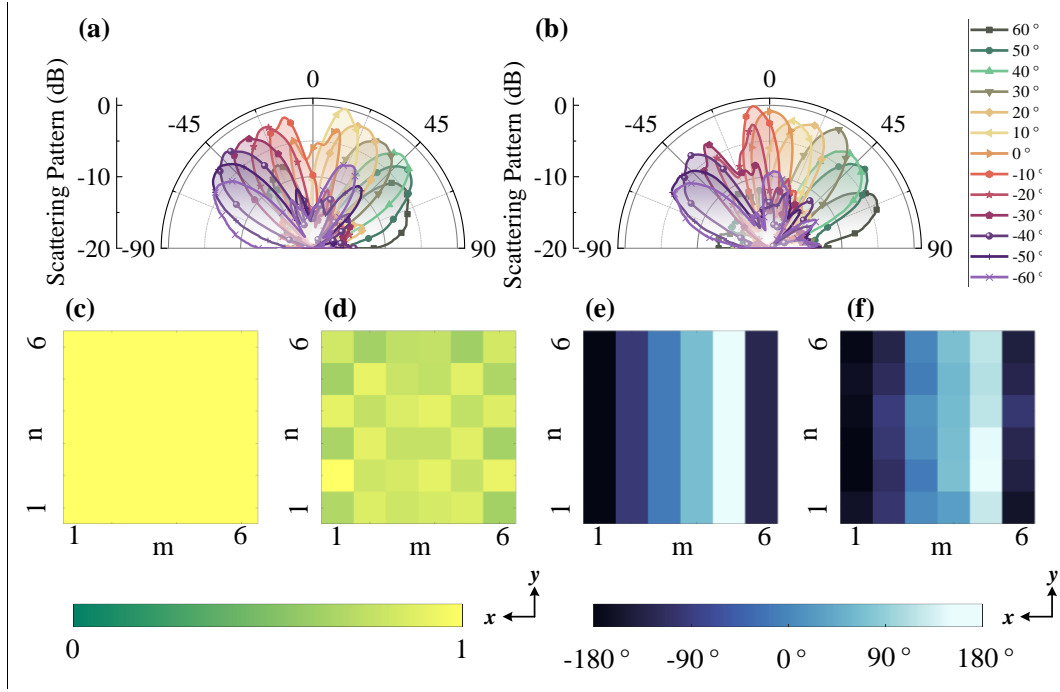

**Fig. 7. (a-b)** Measurement scattering patterns for dynamic beamforming when the feed horn antenna is placed at $\theta = 0°$ and -20°, respectively. **(c, e)** The calculated magnitudes and phases of the received signals by the 6×6 sensing elements with Equation (S1). **(d, f)** The measured magnitudes and phases by the 6×6 sensing elements, where m and n denote the indices of column and row.

Finally, we conduct an inspection of the joint sensing and beam regulation capabilities of the metasurface to determine whether the two can operate independently and concurrently for ISAC applications. **Fig. 2b** illustrates a typical ISAC scenario, which consists of an AMS, the base station, the user ends (transmitter and receiver), and unknown scatterers. The AMS can separately obtain all CSI

$$\boldsymbol{f} = \begin{bmatrix} f_{1,1} & f_{1,2} & \cdots & f_{1,N} \\ f_{2,1} & f_{2,2} & \cdots & f_{2,N} \\ \vdots & \vdots & \ddots & \vdots \\ f_{M,1} & f_{M,2} & \cdots & f_{M,N} \end{bmatrix} \in \mathbb{C}^{M \times N} \quad \text{and} \quad \boldsymbol{g} = \begin{bmatrix} g_{1,1} & g_{1,2} & \cdots & g_{1,N} \\ g_{2,1} & g_{2,2} & \cdots & g_{2,N} \\ \vdots & \vdots & \ddots & \vdots \\ g_{M,1} & g_{M,2} & \cdots & g_{M,N} \end{bmatrix} \in \mathbb{C}^{M \times N} \quad , \quad \text{in}$$

which $f_{m,n}$ and $g_{m,n}$ represent the transmission coefficient between the (m, n)th element and the transmitter and the receiver, respectively. The AMS then adjusts the reflection matrix **Φ** using the adaptive algorithm to optimize communication quality between the transmitter and the receiver (See more details in the Supplementary Information). For experimental validation, three typical ISAC scenarios are considered as illustrated in **Figs. 6b** and **6c**.

In the first scenario (**Fig. 6b**), the signal is transmitted non-line-of-sight between the transmitter and the receiver through the metasurface, while the propagation path is blocked by a triangular metallic obstacle. The surrounding absorbing materials are used to eliminate the multipath effect. Without loss of generality, the transmitter and the receiver are positioned at angles of $\theta$ = -40° and 20°, respectively, in the plane of $\varphi$ = 0°. The phase information of all the components in $f$ and $g$ matrices are measured and depicted in **Figs. 8a** and **8b**. Based on the adaptive algorithm (Equation (S8) in Supplementary Information), the optimal reflection matrix **Φ** is generated, as shown in Fig. **8c**. The transmitter then transmits a QPSK-modulated signal, and the received constellation diagram and the signal spectrum at the receiver are given in **Fig. 9a**. For comparison, a metallic board of the same size as the AMS is placed in the same position instead, and the corresponding received constellation diagram and the signal spectrum are shown in **Fig. 9b**. Additionally, in the same scenario, the AMS is loaded with **Φ** calculated by the traditional beamforming algorithm, and the measurement is repeated, as shown in **Fig. 9c**. Note that, here it is necessary to provide the positions of the transmitter and receiver to AMS in advance (i.e., $\theta$ = -40° and 20°), which reveals one of the drawbacks of the traditional beamforming algorithm. In contrast, the proposed AMS with the adaptive algorithm does not require any prior input of external environmental information.

In contrast to the static metallic board, the AMS can provide an additional channel gain of approximately 19 dB from **Figs. 9a** and **Fig. 9b**. This gain enhancement indicates a substantial improvement of the wireless communication quality provided by the proposed AMS. From **Figs. 9a** and **9c**, it is seen that in complex wireless situations, the AMS exhibits a notable enhancement in adaptability compared to traditional methods, even without prior information on the locations of the transmitter and receiver.

The second scenario **(Fig. 6c)** takes into account an open corridor environment, in which the signals from the base station to the user end undergo distinct multiple effects. The phases of *f* and *g* matrices are measured and depicted in **Figs. 8d** and **8e**, respectively, which display obvious interference to previous cases due to multipath effects and the scattering from obstacles. Here the transmitter also transmits a QPSK-modulated signal. The adaptive algorithm is applied to the AMS to generate the required matrix Φ as shown in **Fig. 8f**. The traditional beamforming algorithm is then applied to the AMS for comparison. Additionally, a metallic board of the same size as the AMS is also used to replace the AMS to observe the change of the received signal quality. The constellation diagram and the signal spectrum received by the receiver under these conditions are shown in **Figs. 9d** to **9f.**

Clearly, the AMS outfitted with an adaptive algorithm can supply an 8dB gain boost in the cascaded channel. However, using the traditional beamforming algorithm the performance of AMS is significantly degraded, which is only comparable to a metallic board. This is because the traditional beamforming algorithm represented by Equation (S4) is based on the assumption of an ideal free-space propagation environment, which may fail in typical open corridor environment like Scenario #2.

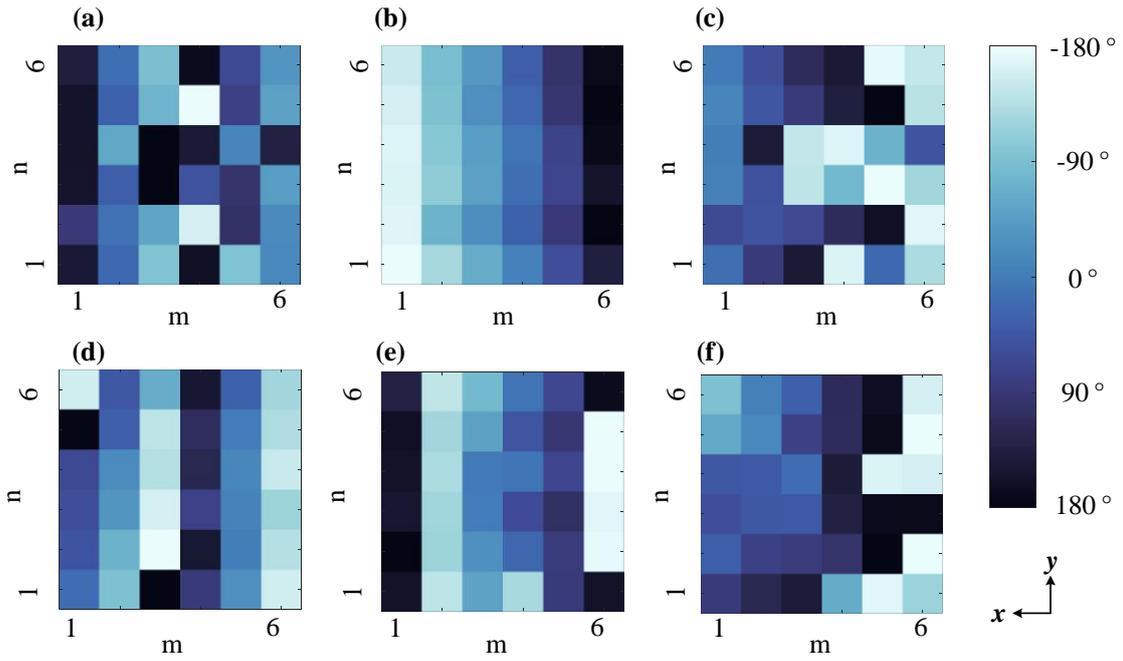

**Fig. 8.** (**a**, **b**) Phases of *f* and *g* measured in Scenario #1. (**c**) The phase of optimal reflection matrix **Φ** in Scenario #1. (**d**, **e**) Phases of *f* and *g* measured in Scenario #2. (**f**) The phase of optimal reflection matrix **Φ** in Scenario #2.

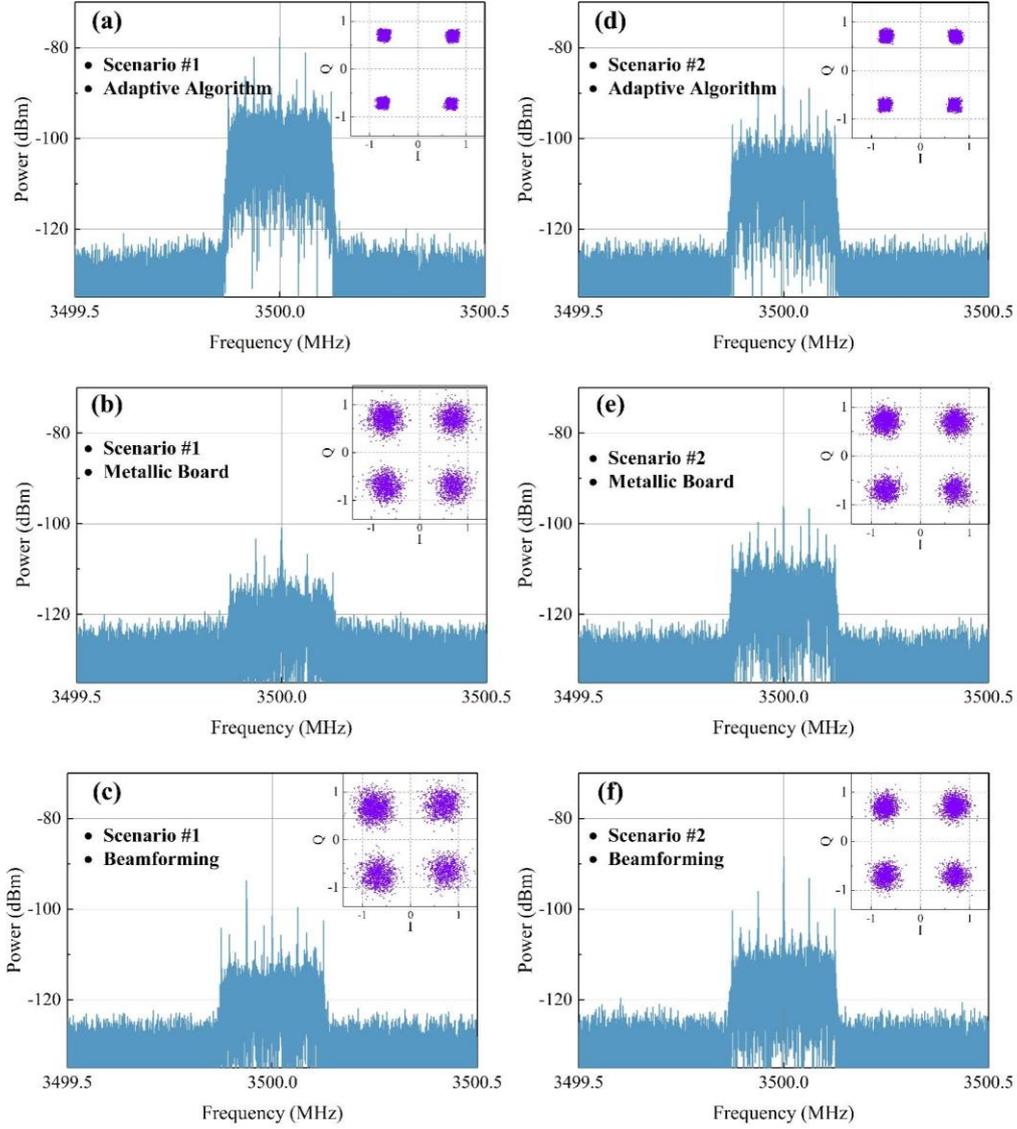

**Fig. 9.** The received constellation diagram and signal spectrum at the receiver under different scenarios. (a, d) Measurement results under Scenario #1 and #2, when the adaptive algorithm described by Equation (S8) is applied to the AMS. (b, e) Measurement results under Scenario #1 and #2, when the AMS is replaced with a metallic board of the same size. (a, d) Measurement results under Scenario #1 and #2, when the traditional beamforming algorithm described by Equation (S4) is applied to the AMS.

Besides the two complex scenarios described above, a third scenario is established as shown in **Fig. 6d,** where the signal is transmitted non-line-of-sight between the transmitter and the receiver through the metasurface without any obstacle in the propagation path. In this case, the AMS can enhance communication quality with traditional beamforming algorithms, similar to conventional metasurfaces, while simultaneously performing DOA estimation to identify the directions of the transmitter and the receiver. The measurement results in this scenario are provided in the Supplementary Information.

In summary, assessing AMS's performance in these scenarios highlights its advantages over conventional metasurfaces: Firstly, AMS exhibits superior adaptability to complex environments thanks to its universal strategy of CSI sensing. Secondly, by taking advantage of its closed-loop working mode of sensing, analysis, and control, the AMS can operate independently without extra information about the channel in advance. Meanwhile, the performances of conventional metasurfaces and algorithms are likely to degrade significantly in complex environments. On the whole, the measured results suffice to demonstrate the significant advantages of AMS in intelligent sensing and manipulating.

**Conclusion**

An AMS with simultaneous manipulating and sensing capabilities is proposed in his paper. The working principle and prototype implementation are demonstrated, and its performances are validated by experimental measurements. The measurement results show that, AMS can manipulate EM waves over a phase range of 0 to 360 degrees while maintaining a reflection loss of no more than 1.45 dB. Additionally, AMS can simultaneously sense the complex amplitude of the incident signal element by element. Three AMS-assisted ISAC scenarios are constructed and experimented. The results reveal that the proposed AMS can significantly enhance communication quality in the complex propagation environments intelligently and provide more powerful capabilities for the in-depth application of metasurfaces in wireless systems in the future.

**Supplementary Information**

**Model of Propagation and Manipulation**

The process of propagation and manipulation of the EM waves shown in **Fig. 2a** can be described as follows. Here the wave originates from the source, traverses through the complex EM environment, and eventually arrive at the AMS. The electric fields $\mathbf{E}_{in}(\mathbf{r}_{m,n})$ incident on the ($m$, $n$)th AMS element can be obtained by

$$\mathbf{E}_{in}(\mathbf{r}_{m,n}) = \int_V \mathbf{M}(\mathbf{r}') \times \nabla G(\mathbf{r}_{m,n}, \mathbf{r}') d\mathbf{r}' - j\omega\mu_0 \int_V \left[1 + \frac{1}{k^2}\nabla\nabla\right] \mathbf{J}(\mathbf{r}') \cdot G(\mathbf{r}_{m,n}, \mathbf{r}') d\mathbf{r}' \quad (S1)$$

in which $\omega$ and $k$ are the angular frequency and wave number, respectively. $\mathbf{r}_{m,n}$ and $\mathbf{r}'$ denote the coordinates of the ($m$, $n$)th AMS elements and the source, respectively. $G(\mathbf{r}_{m,n}, \mathbf{r}')$ is the Green's function of the scalar Helmholtz equation. The incident EM waves induce surface currents, leading to a generation of a secondary radiated field $\mathbf{E}_{out}$. This process is manipulated by the state of the AMS, specifically:

$$\mathbf{E}_{out}(\mathbf{r}_{m,n}) = \Gamma_{m,n} \cdot \mathbf{E}_{in}(\mathbf{r}_{m,n}) = \mathbf{Z} \cdot \mathbf{J}^s(\mathbf{r}_{m,n}) \quad (S2)$$

in which $\Gamma_{m,n} = A_{m,n} e^{j\phi_{m,n}}$ is the complex reflection coefficient of the ($m$, $n$)th element. $\mathbf{J}^s(\mathbf{r}_{m,n})$ is the induced current density excited on the $n$th AMS element and $\mathbf{Z}$ is the matrix that links the sources and the fields. Ultimately, substituting $\mathbf{J}^s$ for $\mathbf{J}$ in Equation (S1), the field $\mathbf{E}$ at any point $\mathbf{r}$ in the space is obtained, as described by Equation (1) in the main text. In Equation (1), $\mathbf{E}_{in}(\mathbf{r})$ is the field directly formed by the waves emitted from the sources, and can be derived by substituting $\mathbf{r}_{m,n}$ with $\mathbf{r}$ in Equation (S1). The remaining component ($\sum_{m=1}^{M}\sum_{n=1}^{N}(-j\omega\mu_0)\left[1+\frac{1}{k^2}\nabla\nabla\right]\cdot\Gamma_{m,n}\cdot\mathbf{Z}^H\cdot\mathbf{E}_{in}(\mathbf{r}_{m,n})\cdot G(\mathbf{r}, \mathbf{r}_{m,n})$) represents the field generated by the waves reflected by the metasurface. The combination of these fields yields the total electric field $\mathbf{E}(\mathbf{r})$ at point $\mathbf{r}$. $H$ stands for conjugate transpose. $G$ is the Green's function. In free space, the Green's function between two points $\mathbf{r}$ and $\mathbf{r}'$ can be expressed as

$$G(\mathbf{r}, \mathbf{r}') = -\frac{e^{-jk|\mathbf{r}-\mathbf{r}'|}}{4\pi|\mathbf{r}-\mathbf{r}'|} \tag{S3}$$

However, ideal free space is generally only achievable under ideal laboratory settings. In realistic scenarios, the propagation of waves is usually unpredictable and the Green's function $G$ is affected by all the media in the environment and has no analytical expression.

**Slot Antennas**

The slot antenna shares the same layered structure with the proposed AMS shown in **Fig. 3a**, with the thickness and substrate information detailed in **Table S1**. The antenna comprises two patches, an H-shaped slot, and a section microstrip line on the back side of the middle layer, as described in the main text. The geometric details are illustrated in **Figs. S1a to S1d**, and its radiation pattern is depicted in **Fig. S1e**. The microstrip line is connected to a feeding port in the current configuration, while in the proposed AMS element, it interfaces with the phase shifter. The metal board (the bottom layer of the AMS) can be regarded as the reflecting surface of the slot antenna and is positioned at a 1/4 wavelength distance from the middle layer. This board can theoretically increase the antenna's gain by 3dB, and enhance the reflection efficiency of AMS, as no energy is radiated into the backside.

**Programmable Phase Shifter**

The phase shifter is based on a 3-dB branch-line coupler, with its details depicted in **Fig. S2a**. The phase of its programmable transmission coefficient is manipulated by tuning the reconfigurable equivalent capacitor $C_{VAR}$ of the varactor diode. And the practical variable capacitance is accomplished by applying different DC voltages in experiments. The transmission phase ranges from -90.2° to 96.7°, encompassing a manipulation range exceeding 180°, as depicted in **Fig. S2b**. In the AMS, the reflected wave traverses the phase shifter twice before being reradiated. Thus, the capability of the proposed design is sufficient for the AMS to manipulate the EM wave across a reflective phase range over 360°. The varactor diode used is MAVR-000120-1411, and **Fig. S2c** illustrates the manufacturer-provided relationship curve between the applied DC voltage and its capacitance.

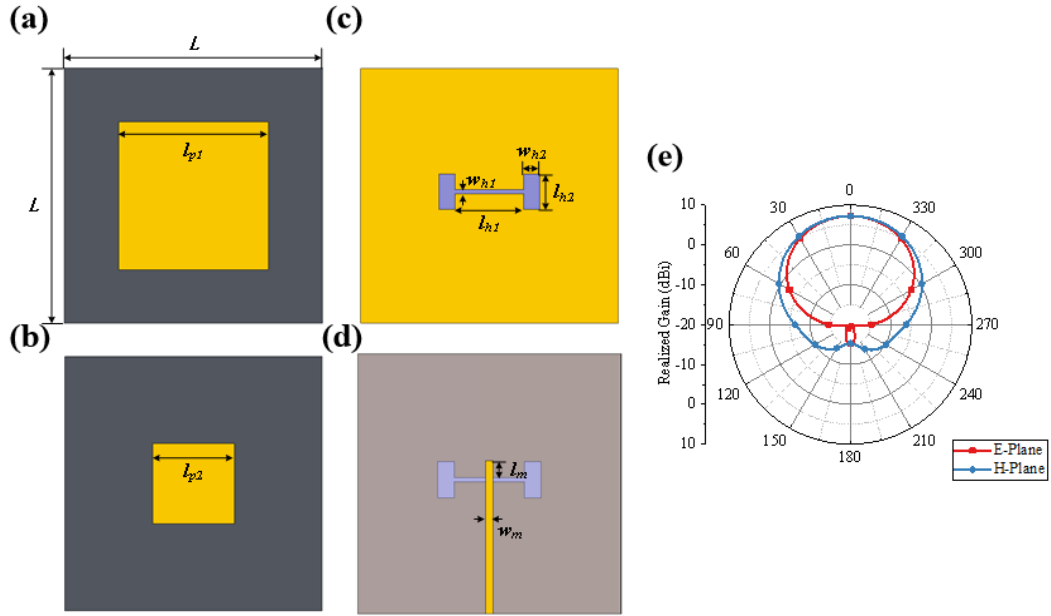

**Fig. S1.** (**a, b**) The front and back side of the top layer of the AMS element, in which $l_{p1}$ = 23.20 mm, $l_{p2}$ = 12.65 mm, $l_{h1}$ = 10.67 mm, $l_{h2}$ = 5.55 mm, $w_{h1}$ = 0.72 mm, $w_{h2}$ = 2.50 mm, $l_m$ = 2.56 mm, $w_m$ = 1.10 mm. (**c, d**) The front and back side of the middle layer of the AMS element, in which the microstrip line is connected to a feeding port instead of phase shifter. (**e**) The radiation patterns of the slot antenna in E-plane and H-plane.

**Table S1. The thickness of each layer of the AMS and the slot antenna.**

| Layer | Thickness |
|---|---|
| F4BTM | 3.00 mm |
| Air (Upper) | 4.00 mm |
| Rogers 4003C | 0.508 mm |
| Air (Lower) | 22.0 mm |
| Copper Board | 0.035 mm |

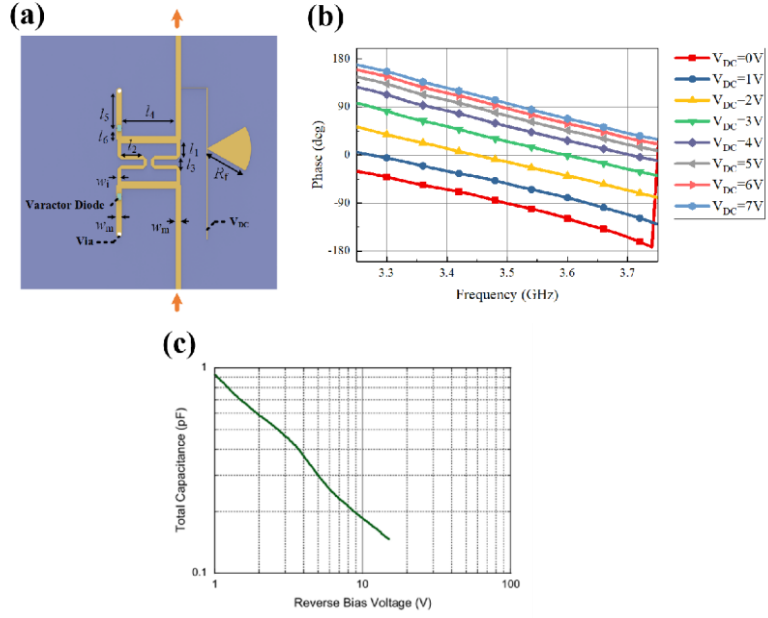

**Fig. S2. (a)** Geometric details of the programmable phase shifter, in which $w_m = 1.10$ mm, $w_i = 0.74$ mm, $l_1 = 3.16$ mm, $l_2 = 4.30$ mm, $l_3 = 2.60$ mm, $l_4 = 10.46$ mm, $l_5 = 7.30$ mm, $l_6 = 2.20$ mm, and $R_f = 8.65$ mm. **(b)** The relationship between the phase shift and the applied DC voltage. **(c)** The relationship between the capacitance of the varactor diode and the applied DC voltage.

**Beamforming Algorithm**

To adjust the beam direction of the scattering pattern, the phases of the reflection coefficients are supposed to be set as

$$\phi_{m,n} = -\phi_{m,n}^{in} - \frac{2\pi}{\lambda}\left(x_{m,n}\sin\theta_0\cos\varphi_0 + y_{m,n}\sin\theta_0\sin\varphi_0\right) \quad \text{(S4)}$$

in which $(x_{m,n}, y_{m,n})$ denotes the coordinates of the $(m, n)$th AMS element, $(\theta_0, \varphi_0)$ represents the beam direction of the scattering pattern, $\lambda$ is the wavelength in free space, $\phi_{m,n}^{in}$ is the phase of the incident wave at the position of the $(m, n)$th element. In particular, when a plane wave illuminates the AMS, we have

$$\phi_{m,n}^{in} = \frac{2\pi}{\lambda}\left(x_{m,n}\sin\theta_0^{in}\cos\varphi_0^{in} + y_{m,n}\sin\theta_0^{in}\sin\varphi_0^{in}\right) \quad \text{(S5)}$$

in which $\left(\theta_0^{in}, \varphi_0^{in}\right)$ represents the direction of the incident wave.

Equations (S4) and (S5) reveal that the algorithm requires the incoming wave's angle of arrival and the desired beamforming direction, namely $\left(\theta_0^{in}, \varphi_0^{in}\right)$ and $(\theta_0, \varphi_0)$, as inputs. This is also why the main text emphasizes that without accurate knowledge of the orientations of the transmitter and the receiver, the beamforming algorithm cannot be executed.

**DOA Estimation Methods**

The simplest method based on the phase difference between the adjacent elements due to incident waves is adopted for DOA estimation. Taking one-dimensional estimation as an example, all AMS elements are uniformly positioned at a distance of $d$ as illustrated in **Fig. S3**, the DOA can be obtained by

$$\theta_0^{in} = \arcsin\left(\frac{\lambda}{2\pi d} \cdot \Delta\phi^{in}\right) \tag{S6}$$

in which $\Delta\phi^{in}$ is the phase difference of the incident wave on the adjacent elements.

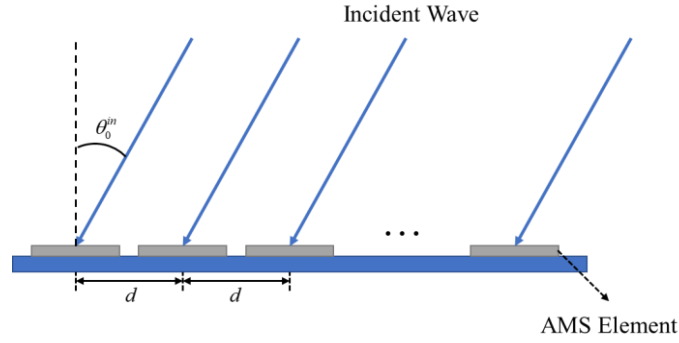

**Fig. S3**. Principle of one-dimensional DOA estimation.

**AMS's method for obtaining CSI in ISAC scenarios**

AMS' sensing process to obtain CSI is summarized as follows: Each communication terminal (transmitter and receiver) sequentially transmits a pilot signal to the metasurface. By acquiring the field distribution $\mathbf{E}_{in}(\mathbf{r}_{m,n})$ on its surface twice, the AMS can separately obtain all CSI $\mathbf{f} = \begin{bmatrix} f_{1,1} & f_{1,2} & \cdots & f_{1,N} \\ f_{2,1} & f_{2,2} & \cdots & f_{2,N} \\ \vdots & \vdots & \ddots & \vdots \\ f_{M,1} & f_{M,2} & \cdots & f_{M,N} \end{bmatrix} \in \mathbb{C}^{M\times N}$ and

$\mathbf{g} = \begin{bmatrix} g_{1,1} & g_{1,2} & \cdots & g_{1,N} \\ g_{2,1} & g_{2,2} & \cdots & g_{2,N} \\ \vdots & \vdots & \ddots & \vdots \\ g_{M,1} & g_{M,2} & \cdots & g_{M,N} \end{bmatrix} \in \mathbb{C}^{M\times N}$. $f_{m,n}$ and $g_{m,n}$ represent the transmission coefficient between the ($m,n$)th element and the transmitter and the receiver, respectively, which are proportional to $\mathbf{E}_{in}(\mathbf{r}_{m,n})$ when the transmitter and the receiver transmit the pilot signal.

In simple wireless environments, characterized by ideal free space without scatterers, all transmissions occur under line-of-sight (LoS) conditions and the Green's function can be calculated by Equation (S3). With these assumptions, the phases of both *f* and *g* are uniformly varying along the *x*-axis and *y*-axis, displaying a gradient distribution. This uniform variation ensures the feasibility of DOA estimation as given by Equations (S6).

In complex wireless environments, the values of *f* and *g* become unpredictable and no longer follow a regular pattern. This may cause a decline in the accuracy of traditional DOA estimation and beamforming algorithms, or even render them ineffective. This reveals the significant impact of the propagation environment on the performance of wireless systems, highlighting the importance of the proposed AMS's capability for element-by-element sensing.

**Adaptive algorithm for adjusting the reflection matrix in ISAC scenarios**

The adaptive algorithm aims to find the optimal reflection matrix $\mathbf{\Phi}$ to enhance communication quality between terminals, namely, it seeks to maximize the electric field magnitude $|\mathbf{E}(\mathbf{r})|$ at the receiver's location when the transmitter serves as the source. Considering the direct path between the transmitter and the receiver is ineffective (indicated by a cross in **Fig. 2b**), the optimization goal can be formulated as

$$\max_{\mathbf{\Phi}} \ \left| \mathbf{g}' \cdot \mathrm{diag}(\mathbf{\Phi}') \cdot \mathbf{f}'^{\mathrm{T}} \right| \quad\quad\quad (S7)$$
$$\mathrm{s.t.} \ \ |\Gamma_{m,n}| = A_{m,n} \leq 1$$

in which $\mathbf{g}' = \mathrm{vec}(\mathbf{g})^T = [g_{1,1}, g_{1,2}, \cdots g_{1,N}, g_{2,1}, g_{2,2}, \cdots g_{2,N}, \cdots, g_{M,1}, \ldots, g_{M,N}] \in \mathbb{C}^{MN\times 1}$,
$\mathbf{\Phi}' = \mathrm{vec}(\mathbf{\Phi})^T = [\Gamma_{1,1}, \Gamma_{1,2}, \cdots \Gamma_{1,N}, \Gamma_{2,1}, \Gamma_{2,2}, \cdots \Gamma_{2,N}, \cdots, \Gamma_{M,1}, \ldots, \Gamma_{M,N}] \in \mathbb{C}^{MN\times 1}$,
$\mathbf{f}' = \mathrm{vec}(\mathbf{f})^T = [f_{1,1}, f_{1,2}, \cdots f_{1,N}, f_{2,1}, f_{2,2}, \cdots f_{2,N}, \cdots, f_{M,1}, f_{M,2}, \ldots, f_{M,N}] \in \mathbb{C}^{MN\times 1}$,
$\left| \mathbf{g}' \cdot \mathrm{diag}(\mathbf{\Phi}') \cdot \mathbf{f}'^{\mathrm{T}} \right|$ is the cascaded channel gain between the transmitter and the receiver.

The AMS then adaptively adjusts $\Phi$ according to the following equation:

$$\phi_{m,n} = \angle\Gamma_{m,n} = -\left(\angle f_{m,n} + \angle g_{m,n}\right) \tag{S8}$$

in which $\phi_{m,n} = \angle\Gamma_{m,n}$ is the reflection phase of the $(m,n)$th element. The adaptive algorithm represented by Equation (S8) provides the optimal solution to the problem posed by Equation (S7) (considering all the reflection magnitudes are approximately 1), as it ensures that all signals reaching the receiver are in phase.

It should be noted that the adaptive algorithm, when applied in free space, is equivalent to the traditional beamforming algorithm represented by Equation (S4). Therefore, the adaptive algorithm has a wider range of applicability, and the traditional beamforming algorithm is just a special case of it.

**Measurement results in the experiment of ISAC Scenario #3**

Scenario #3 (**Fig. 6d**) represents the simplest communication environment, where both links of transmitter-AMS and AMS-receiver maintain a LoS configuration. The transmitter and the receiver are positioned at angles of $\theta$ = -40° and 20°, respectively, in the plane of $\varphi$ = 0°. The phase information of all components of $f$ and $g$ is measured and depicted in **Figs. S4a** and **S4b**. Subsequently, the matrix $\Phi$ is correspondingly adjusted according to the traditional beamforming algorithm (Equation (S4)), as shown in **Fig.S4c** and the transmitter transmits a QPSK-modulated signal. The received constellation diagram and the signal spectrum at the receiver are shown in **Fig. S4d**. Following this, a metallic board with an equal size to the AMS is placed at the same position instead, and the received constellation diagram and the signal spectrum are measured again, as shown in **Fig. S4e** for comparison.

As illustrated in **Figs. S4d** and **S4e**, compared to a static metallic board, AMS can provide an additional channel gain of approximately 16 dB with its beamforming capability. Additionally, the phases of $f$ and $g$ vary nearly uniformly in the horizontal direction, facilitating DOA estimation: according to Equation (S6), the transmitter and the receiver are located in the directions of $\theta$ = -41.3° and $\theta$ = 22.7°, respectively.

The experimental results from Scenario #3 indicate that in the nearly ideal propagation environment set up in the laboratory, the proposed AMS effectively performs the task of improving communication quality through directional beamforming, similar to conventional metasurfaces. Additionally, it is worth noting that the AMS does not require advance knowledge and can autonomously identify the positions of the transmitter and receiver.

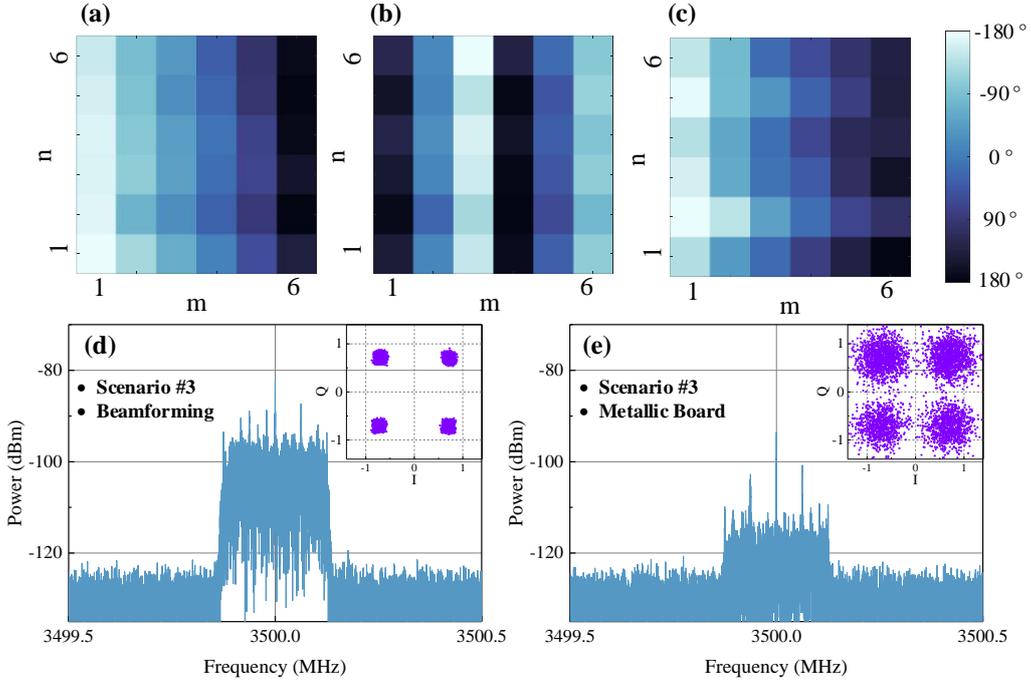

Fig. S4. (**a**, **b**) Phases of *f* and *g* measured in Scenario #3. (**c**) Optimized reflection matrix **Φ** in Scenario #3. (**d**, **e**) The received constellation diagram and the signal spectrum at the receiver when the AMS and metallic board is placed.

**RF switch matrix**

As shown in Fig. S5, the RF switch matrix is connected to all 6×6 sensing chains via cables. This enables the wireless receiver system to capture the coupled signals from all AMS elements using fewer channels. During the measurements, the switch matrix is electronically controlled to quickly toggle between selected links, allowing for rapid acquisition of information from all 36 signal paths.

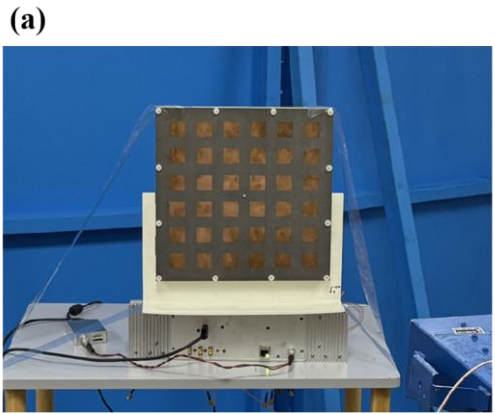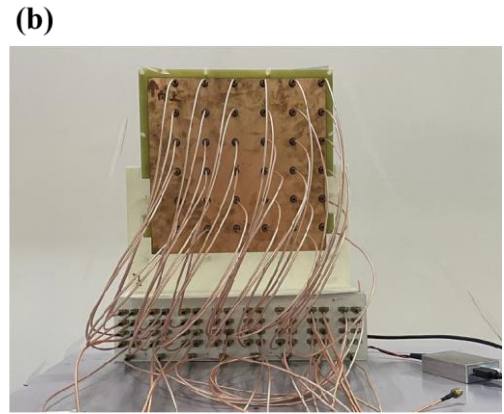

Fig. S5. (**a**, **b**) The front and rear perspectives of the RF switch matrix and AMS.